	\documentclass[12pt, draftclsnofoot, onecolumn]{IEEEtran}
	
	\usepackage{pifont,bm,multicol,amsfonts,amsmath,color,amssymb,graphicx, epsfig,cite,psfrag,subfigure,algorithm,enumerate,stfloats,algorithmic,epstopdf,balance,multirow,multicol,makecell}

\begin{document}

\title{Deep Learning Based RIS Channel Extrapolation with Element-grouping}

\author{Shunbo Zhang, Shun Zhang, \emph{Senior Member, IEEE}, Feifei Gao, \emph{Fellow, IEEE}, Jianpeng Ma, \emph{Member, IEEE}, Octavia A. Dobre, \emph{Fellow, IEEE}
    \thanks{S. Zhang, S. Zhang and J. Ma are with the State Key Laboratory of Integrated Services Networks, Xidian University, Xi'an 710071, P. R. China (e-mail: sbzhang$\_$19@stu.xidian.edu.cn, zhangshunsdu@xidian.edu.cn, jpmaxdu@gmail.com).

    F. Gao is with Institute for Artificial Intelligence, Tsinghua University (THUAI),
    State Key Lab of Intelligent Technologies and Systems, Tsinghua University,
    Beijing National Research Center for Information Science and Technology (BNRist),
    Department of Automation,Tsinghua University
    Beijing, P.R. China (email: feifeigao@ieee.org).

    O. A. Dobre is with Faculty of Engineering and Applied Science, Memorial University, St. John's NL AIC-5S7, Canada (e-mail: odobre@mun.ca).}
}

\maketitle

\vspace{-10mm}
\begin{abstract}
	Reconfigurable intelligent surface (RIS) is considered as a revolutionary technology for future wireless communication networks.
	In this letter, we consider the acquisition of the cascaded channels, which is a challenging task due to the massive number of passive RIS elements.
	To reduce the pilot overhead, we adopt the element-grouping strategy, where each element in one group shares the same reflection coefficient and is assumed to have the same channel condition.
	We analyze the channel interference caused by the element-grouping strategy and further design two deep learning based networks.
	The first one aims to refine the partial channels by eliminating the interference, while the second one tries to extrapolate the full channels from the refined partial channels.
	We cascade the two networks and jointly train them.
	Simulation results show that the proposed scheme provides significant gain compared to the conventional element-grouping method without interference elimination.

\end{abstract}

\maketitle
\thispagestyle{empty}
\vspace{-1mm}


\section{Introduction}
\label{introduce}

The increasingly demanding objectives for the sixth generation (6G) wireless communication networks have spurred recent research activities on novel wireless hardware architectures \cite{Huang_RIS}.
As an artificial planar structure with integrated electronic circuits, the reconfigurable intelligent surface (RIS) can be programmed to manipulate an incoming electromagnetic field in a wide variety of functionalities \cite{Tang_RIS}.
The amplitude and phase information at each RIS element can be adjusted independently, so as to digitally manipulate the electromagnetic waves and realize signal propagation direction regulation and in-phase superposition in the three-dimensional space, which will improve the quality of received signals and the performance of wireless communication \cite{RIS2}.
Hence, incorporating RISs in wireless networks has been advocated as a revolutionary means to transform any naturally passive wireless communication environment, including the set of objects between a transmitter and a receiver,  to an active one, and has become a hot research topic \cite{Huang_RIS, Tang_RIS, RIS2, Octavia_RIS}.



To achieve good beamforming performance, RIS usually contains a large number of elements.
However, the more elements are used, the more unknown parameters exist, which inevitably increases the overhead and complexity of channel estimation.
To tackle this problem, Yang \emph{et al.} proposed an element-grouping strategy \cite{element_grouping1}.
In the element-grouping strategy, every RIS element in each group is in the open state, shares the same reflection coefficient and is assumed to have the same channel state information, which greatly reduces the pilot overhead and enhances the system's spectrum efficiency \cite{element_grouping2}.
In \cite{element_grouping3}, Zheng \emph{et al.} implemented two efficient channel estimation schemes with the element-grouping strategy.
However, the channels related with different RIS elements cannot be ensured to be identical, and only partial channel information can be obtained within the element-grouping scheme.

Fortunately, for a given RIS structure, there exists deterministic mapping between the partial and the full channel information.
As mentioned in \cite{Liu_WC}, we can exploit the universal approximation capability of neural networks (NNs), adopt deep leaning (DL) to characterize the above mapping, and extrapolate the full RIS channel information from the partial one.
In fact, DL has been utilized for channel estimation over RIS-assisted systems.
In \cite{DL1}, Khan \emph{et al.} applied DL methods to estimate channels and phase angles in RIS-assisted wireless communication systems.
The trained DeepRIS model is deployed to estimate channels and phase angles from the received symbols.
In \cite{DL2}, Elbir \emph{et al.} proposed a federated learning (FL) framework for channel estimation and designed a convolutional NN (CNN) trained on the local datasets of the users without sending them to the base station (BS).
With respect to the element-grouping scheme, different channels related with distinguished RIS elements of each group would interfere with each other.
The larger the difference among the channels related with different RIS elements in a group is, the stronger the interference is.
This observation should be carefully considered within the design of the DL scheme.

In this letter, we consider the acquisition of the cascaded channels in a RIS-assisted communication system and adopt the element-grouping strategy to reduce the pilot overhead.
Specifically, we investigate the impact of the channel interference among distinguished RIS elements.
We first utilize the pilot orthogonality to eliminate the interference among different element groups.
After that, we construct a DL-based network to eliminate the interference within each element group and acquire the refined partial cascaded channels.
With these partial channels, we further adopt an NN to extrapolate the full cascaded channels.
Lastly, we design a joint network training scheme to optimize the two networks.

\emph{Notations:}
We use lowercase (uppercase) boldface to denote vector (matrix).
$(\cdot )^T$, $(\cdot )^H$, $\|\cdot\|_F$ and $\mathbb{E}\{\cdot\}$ represent the transpose, Hermitian, Frobenius norm and expectation, respectively.
$\otimes$ is the Kronecker product operator and $|\mathcal A | $ is the number of elements in set $\mathcal A$.
$v \sim \mathcal{CN} (0, \sigma^2)$ means that $v$ follows the complex Gaussian distribution with zero-mean and variance $\sigma^2$.
$\mathbf{I}_N$ and $\mathbf{1}_N$ represent a $N\times N$ identity matrix and a $N\times 1$ all-one vector, respectively.
$\text{diag} \{\mathbf x\}$ is a diagonal matrix whose diagonal elements are formed with the elements of $\mathbf x$.

\section{System Model}
\label{model}

\begin{figure}
	\centering
	\includegraphics[width=110mm]{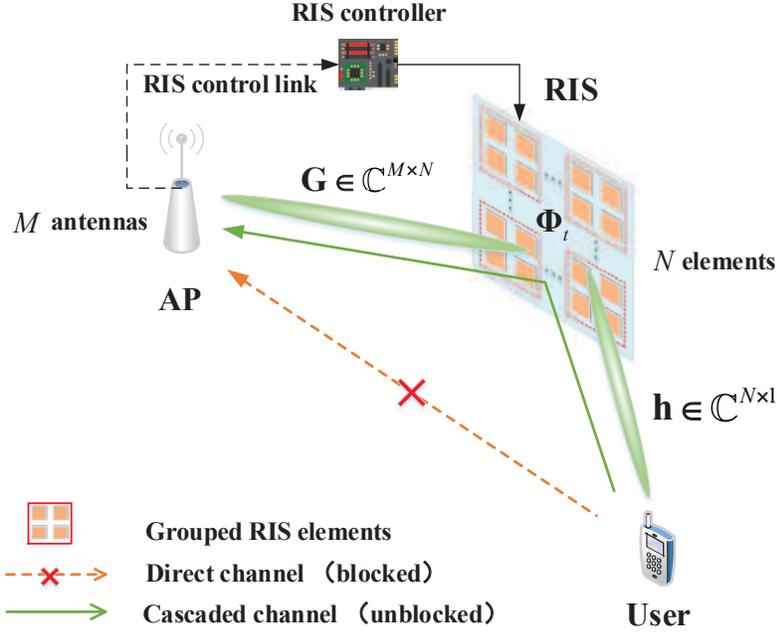}
	\caption{The RIS-assisted uplink communication system.}
	\label{scene}
\end{figure}

As shown in Fig. \ref{scene}, we consider an RIS-assisted communication system where a single antenna user tries to communicate with the access point (AP) via the RIS reflection.
The AP is equipped with $M$ antennas in the form of a uniform linear array (ULA) and RIS consists of $N_v\times N_h=N$ passive reflection elements in the form of a uniform planar array (UPA), where $N_h$ and $N_v$ are the number of elements on the vertical and horizontal dimensions, respectively.
Moreover, RIS is connected to a dedicated controller for dynamically adjusting the individual reflection of each element.
Without loss of generality, in this letter we focus on the uplink communication.
Since the direct link between the user and AP tends to be blocked by possible obstacles like buildings and human bodies, we only consider the RIS-assisted link and ignore the direct link.
The $M\times N$ uplink channel matrix $\mathbf G=[\mathbf g_1,\mathbf g_2,\ldots, \mathbf g_N]$ from RIS to AP is expressed as
\begin{align}
\label{RB_channel}
	\mathbf{G}=\sum_{p=1}^{P_g}\alpha_{g,p}\mathbf{a}_{A}(\theta_{g,p})\mathbf{a}_{R}^H(\vartheta_{g,p},\varphi_{g,p}),
\end{align}
where $\mathbf g_n$ denotes the channel between the $n$-th RIS element and AP, $P_g$ is the number of the scattering path from the RIS to AP, $\alpha_{g,p}$ is the equivalent complex channel gain of the $p$-th path, and $\mathbf{a}_A(\theta_{g,p})\!\in\!\mathbb{C}^{M\times 1}$ and $\mathbf{a}_R(\vartheta_{g,p},\varphi_{g,p})\!\in\!\mathbb{C}^{N\times 1}$ respectively denote the spatial steering vector of AP and  RIS, with $\theta_{g,p}$, $\vartheta_{g,p}$ and $\varphi_{g,p}$ as the angle of arrival (AoA) at AP, the azimuth and the elevation angle of departure (AoD) at the RIS, separately.
$\mathbf{a}_A(\theta_{g,p})\!=\![1,e^{\jmath\frac{2\pi d}{\lambda}\sin\theta_{g,p}},\cdots,e^{\jmath\frac{2\pi d}{\lambda}(M-1)\sin\theta_{g,p}}]^T$, where $d$ is the antenna spacing and $\lambda$ denotes the carrier wavelength.
$\mathbf{a}_{R}(\vartheta_{g,p},\varphi_{g,p})=\mathbf{a}_{el}(\varphi_{g,p})\otimes\mathbf{a}_{az}(\vartheta_{g,p},\varphi_{g,p})$,
where $\mathbf{a}_{el}(\varphi_{g,p})=[1,e^{\jmath\frac{2\pi d}{\lambda}\cos\varphi_{g,p}},\cdots,e^{\jmath\frac{2\pi d}{\lambda}(N_v-1)\cos\varphi_{g,p}}]^T$ is the steering vector in the vertical direction and $\mathbf{a}_{az}(\vartheta_{g,p},\varphi_{g,p})=[1,e^{\jmath\frac{2\pi d}{\lambda}\sin\varphi_{g,p}\cos\vartheta_{g,p}},\cdots,e^{\jmath\frac{2\pi d}{\lambda}(N_h-1)\sin\varphi_{g,p}\cos\vartheta_{g,p}}]^T$ is the steering vector in the horizontal direction.
The $N\times 1$ uplink channel $\mathbf{h}=[h_1,h_2,\cdots,h_N]^T$ from user to RIS is
\begin{align}
	\mathbf{h}=\sum_{p=1}^{P_h}\alpha_{h,p}\mathbf{a}_{R}(\vartheta_{h,p},\varphi_{h,p}),
\end{align}
where $h_n$ is the channel between the $n$-th RIS element and the user, $P_h$ is the number of the scattering path from the user to the RIS, $\alpha_{h,p}$ is the equivalent complex channel gain of the $p$-th path, and $\mathbf{a}_{R}(\vartheta_{h,p},\varphi_{h,p})\in\mathbb{C}^{N\times 1}$ has a similar definition as $\mathbf{a}_{R}(\vartheta_{g,p},\varphi_{g,p})$.
$\vartheta_{h,p}$ and $\varphi_{h,p}$ represent the azimuth and the elevation AoA for RIS, respectively.

To estimate the uplink channels, the user transmits a pilot symbol $s_t$ to AP in the $t$-th time slot and RIS dynamically changes the phase shift of each element in different time slots.
The $M\times 1$ received signal is expressed as
\begin{align}
\label{received_1}
	\mathbf{y}_t=\sqrt{P}\mathbf{G}\bm{\Phi}_t\mathbf{h}s_t+\mathbf{n}_t,
\end{align}
where $P$ is the user transmit power, $\bm{\Phi}_t=\text{diag}\{e^{\jmath\theta_{1,t}}, e^{\jmath\theta_{2,t}}, \cdots, e^{\jmath\theta_{N,t}}\}\in\mathbb{C}^{N\times N}$ denotes the reflection pattern of the RIS in the $t$-th time slot, $\theta_{n,t}\in[-\pi,\pi)$ is the phase shift, and $\mathbf{n}_t\sim\mathcal{CN}(0,\sigma^2\mathbf{I}_M)$ represents the $M\times 1$ additive white Gaussian noise (AWGN) vector.
The signal-to-noise (SNR) is defined as $\text{SNR}=P/\sigma^2$.
Correspondingly, the received signal \eqref{received_1} can be rewritten as
\begin{align}
\label{received_2}
	\mathbf{y}_t&=\sqrt{P}\sum_{n=1}^N\mathbf{g}_nh_ns_te^{\jmath\theta_{n,t}}+\mathbf{n}_t
	=\sqrt{P}\sum_{n=1}^N\mathbf{a}_nx_{n,t}+\mathbf{n}_t
	=\sqrt{P}\mathbf{A}\mathbf{x}_t+\mathbf{n}_t,
\end{align}
where $\mathbf{A}=[\mathbf{a}_1,\mathbf{a}_2,\cdots,\mathbf{a}_N]=[\mathbf{g}_1h_1,\mathbf{g}_2h_2,\cdots,\mathbf{g}_Nh_N]\in\mathbb{C}^{M\times N}$ is the cascaded channel matrix and $\mathbf{x}_t=[x_{1,t},x_{2,t},\cdots,x_{N,t}]^T=[s_te^{\jmath\theta_{1,t}},s_te^{\jmath\theta_{2,t}},\cdots,s_te^{\jmath\theta_{N,t}}]^T\in\mathbb{C}^{N\times 1}$ is the equivalent pilot signal in the $t$-th time slot.
In this letter, we focus on the acquisition of the cascaded channel matrix $\mathbf A$.

\section{DL-based Channel Extrapolation Scheme}

\subsection{Element-grouping Strategy}
\label{element_grouping}

Theoretically, we can design an $N\times T$ channel training matrix $\mathbf{X}=[\mathbf{x}_1,\mathbf{x}_2,\cdots,\mathbf{x}_T]$ that satisfies $\mathbf{X}\mathbf{X}^H=\mathbf{I}_N$ to directly estimate $\mathbf{A}$ with the linear estimator, where $T$ is the time duration of the pilot signal.
Within the Bayesian framework \cite{Ma_TCOM}, $T\geq N$ must be satisfied to effectively recover $\mathbf{A}$, which will cause large pilot overhead.
To overcome this bottleneck, we adopt the element-grouping strategy on RIS and divide $N$ RIS elements into $\widetilde{N}$ groups, as depicted in Fig. \ref{scene}.
In each group, $K=N/\widetilde{N}$ adjacent elements share a common phase shift and are assumed to possess the same channel condition.
The equivalent pilot signal in the $t$-th time slot can be expressed as $\mathbf{x}_t=\tilde{\mathbf{x}}_t\otimes\mathbf{1}_{K}$, where $\tilde{\mathbf{x}}_t=[\tilde{x}_{1,t},\tilde{x}_{2,t},\cdots,\tilde{x}_{\widetilde{N},t}]^T=[s_te^{\jmath\tilde{\theta}_{1,t}},s_te^{\jmath\tilde{\theta}_{2,t}},\cdots,s_te^{\jmath\tilde{\theta}_{\widetilde{N},t}}]^T$ and $\tilde{\theta}_{n,t}$ is the common phase shift of the $n$-th element group.
The grouped cascaded channel matrix can be denoted as $\bar{\mathbf{A}}=\widetilde{\mathbf{A}}\otimes\mathbf{1}_{K}^T$, where $\widetilde{\mathbf{A}}=[\tilde{\mathbf{a}}_1,\tilde{\mathbf{a}}_2,\cdots,\tilde{\mathbf{a}}_{\widetilde{N}}]\in\mathbb{C}^{M\times \widetilde{N}}$ and the $n$-th column in $\widetilde{\mathbf{A}}$ is equal to the $((n-1)K+1)$-th column in $\mathbf{A}$.
Then, to estimate $\bar{\mathbf{A}}$, the minimum time duration of the pilot signal can be reduced from $N$ to $\widetilde{N}$.

After some tedious operations, we can rewrite \eqref{received_2} as
\begin{align}
\label{received_3}
	\mathbf{y}_t
	&=K\sqrt{P}\sum_{n=1}^{\widetilde{N}}\widetilde{\mathbf{a}}_n\widetilde{x}_{n,t}+\sqrt{P}\sum_{n=1}^{\widetilde{N}}\left(\mathbf{a}_{(n-1)K+2}-\widetilde{\mathbf{a}}_n\right)\widetilde{x}_{n,t}\notag\\ &+\cdots+\sqrt{P}\sum_{n=1}^{\widetilde{N}}\left(\mathbf{a}_{(n-1)K+K}-\widetilde{\mathbf{a}}_n\right)\widetilde{x}_{n,t}+\mathbf{n}_t\notag\\
	&=K\sqrt{P}\widetilde{\mathbf{A}}\widetilde{\mathbf{x}}_t+
    \sqrt{P}\underbrace{(\mathbf{A}-\bar{\mathbf{A}})}_{\mathbf V}\mathbf{x}_t+\mathbf{n}_t,
\end{align}
where $\mathbf{V}=\mathbf{A}-\bar{\mathbf{A}}$ represents the channel interference matrix.


As shown in \eqref{received_3}, the information of the entire cascaded channel $\mathbf A$ is incorporated in $\widetilde{\mathbf{A}}$ and $\mathbf{V}$, which are
coupled with each other.
The interference term $\mathbf{V}$ contains both the inter-group and intra-group interferences.
Let $\widetilde{\mathbf{X}}=[\tilde{\mathbf{x}}_1,\tilde{\mathbf{x}}_2,\cdots,\tilde{\mathbf{x}}_T]$ be the pilot matrix.
By leveraging the orthogonality of $\widetilde{\mathbf{X}}$, i.e., $\widetilde{\mathbf{X}}\widetilde{\mathbf{X}}^H=\mathbf{I}_{\widetilde{N}}$, the inter-group interference can be eliminated.
However, the intra-group interference cannot be ignored and would severely impact the accuracy of the channel acquisition.
In this letter, we propose two DL-based networks and jointly train them to eliminate the interference and extrapolate the cascaded channel.

\subsection{DL-based Interference Elimination and Channel Extrapolation}
\label{channel_estimation}

\begin{figure}
	\centering
	\includegraphics[width=110mm]{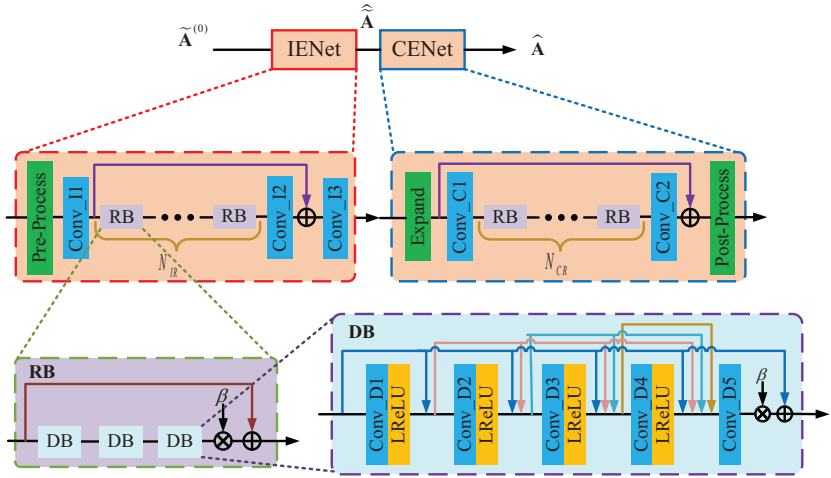}
	\caption{The flowchart and architecture of the designed network.}
	\label{net_flow}
\end{figure}

The flowchart and the network architectures are depicted in Fig. \ref{net_flow}.
In this letter, both the interference elimination and the channel extrapolation are similar to the super-resolution in the field of image processing \cite{Wang_ESRGAN}, which try to enhance the visual quality and recover the complete information from the partial information.

With the perfect knowledge about $\mathbf V$ and $\mathbf{N}=[\mathbf{n}_1,\mathbf{n}_2,\cdots,\mathbf{n}_T]$ , $\widetilde{\mathbf{A}}$ can be written as
\begin{align}
\label{ls_estimation}
	\widetilde{\mathbf{A}}
	=\frac{1}{K\sqrt{P}}(\mathbf{Y}\widetilde{\mathbf{X}}^H-\sqrt{P}\mathbf{V}(\mathbf{I}_{\widetilde{N}}\otimes \mathbf{1}_K)-\mathbf{N}\widetilde{\mathbf{X}}^H),
\end{align}
where $\mathbf{Y}=[\mathbf{y}_1,\mathbf{y}_2,\cdots,\mathbf{y}_T]$ and $\mathbf{X}=[\mathbf{x}_1,\mathbf{x}_2,\cdots,\mathbf{x}_T]=\widetilde{\mathbf{X}}\otimes\mathbf{1}_K$.
In practice, due to the unknown interference term $\mathbf{V}$ and noise term $\mathbf{N}$ in \eqref{ls_estimation}, $\widetilde{\mathbf{A}}$ can not be obtained accurately.
Inspired by the outstanding learning ability of NNs, we
propose a DL-based \textbf{I}nterference \textbf{E}limination \textbf{Net}work (IENet) to refine the least square (LS) estimation of $\widetilde{\mathbf{A}}$, i.e.,
$\widetilde{\mathbf{A}}^{(0)}=\mathbf{Y}\widetilde{\mathbf{X}}^H/(K\sqrt{P})$, through weakening the effect of both  $\mathbf{V}$ and $\mathbf{N}$ as
\begin{align}
\label{elimination}
	\widehat{\widetilde{\mathbf{A}}}=f(\widetilde{\mathbf{A}}^{(0)};\bm{\Theta}),
\end{align}
where $f(\cdot)$ denotes the refinement function learned by IENet,
$\widehat{\widetilde{\mathbf{A}}}$ is the output that represents the refined version of $\widetilde{\mathbf{A}}$ and $\bm{\Theta}$ contains the tunable parameters of IENet.

As shown in Fig. \ref{net_flow}, since the DL model can only deal with real values, we add a pre-processing module to the head of IENet to first separate the real and imaginary parts of $\widetilde{\mathbf{A}}^{(0)}$ and stack them up on a new dimension.
Thus, the input dimension of IENet is transformed from $M\times \widetilde{N}$ to $2\times M\times \widetilde{N}$.
We construct IENet with 3 convolutional layers (``Conv\_I1''-``Conv\_I3'' in Fig. \ref{net_flow}) and $N_{IR}$ cascaded residual blocks (RBs) that have the same structure.
Each RB contains 3 identical dense blocks (DBs); each DB is composed of 5 convolutional layers (``Conv\_D1''-``Conv\_D5'' in Fig. \ref{net_flow}) and the \emph{Leaky ReLU}\footnote{Leaky ReLU is an improvement of ReLU; when its input $x\geq 0$, its output is $y=x$, and when $x< 0$, its output is $y=\alpha x$, where $\alpha\in(0,1)$} (LReLU) is adopted as the activation function for the first 4 layers in one DB.
Conv\_I1 applies 32 kernels with size 5$\times$5 to extract the features of the input with a bigger receptive field.
All the subsequent convolutional layers in IENet apply 32 kernels with size 3$\times$3 except Conv\_I3, which applies 2 kernels with size 3$\times$3 to output the refined $\widehat{\widetilde{\mathbf{A}}}$.
Inspired by the residual network (ResNet)\cite{He_ResNet}, we add many shortcuts in the designed networks to avoid the \emph{gradient vanishing}, \emph{gradient exploding} and \emph{network degradation} in training a deep NN.
For example, within both RB and DB, the result of the \emph{residual} mapping, i.e., the result learned by the stacked nonlinear transformations, is scaled down by multiplying a constant $\beta$ and then added to the result of the \emph{identity} mapping, i.e., the input of RB/DB, for a stable network training.
To further enhance the network capacity, within one DB, the feature maps of all previous layers are empirically concatenated as the input of each subsequent convolutional layer, which is referred to as the dense connection.

Since the reflection elements of RIS are placed close to each other, the channels at different elements are highly correlated and it is possible to infer  $\mathbf{A}$ from $\widehat{\widetilde{\mathbf{A}}}$.
Therefore, we construct a DL-based \textbf{C}hannel \textbf{E}xtrapolation \textbf{Net}work (CENet) to extrapolate the full cascaded channel $\mathbf{A}$ as
\begin{align}
\label{extrapolation}
	\widehat{\mathbf{A}}=g(\widehat{\widetilde{\mathbf{A}}};\bm{\Xi}),
\end{align}
where $g(\cdot)$ denotes the mapping function learned by CENet, $\widehat{\mathbf{A}}$ is the output that represents the extrapolated version of $\mathbf{A}$, and $\bm{\Xi}$ contains the tunable parameters of CENet.



\begin{table}[!t]
	\centering
	\renewcommand{\arraystretch}{1.3 }
	\caption{Structures of IENet and CENet.}
	\label{table_Net_structure}
	\begin{tabular}{c|c|c|c|c}
		\hline
		&\makecell[c]{Layer \\ name} &\makecell[c]{Output \\ size}&\makecell[c]{Kernel \\ size} &\makecell[c]{Activation \\ function}\\
        \hline
		\multirow{4}*{IENet}
		&Conv\_I1 &$32\times M\times \widetilde{N}$ &$5\times 5$ &-\\
		\cline{2-5}
		&Conv\_D1-Conv\_D4 &$32\times M\times \widetilde{N}$ &$3\times 3$ &LReLU\\
		\cline{2-5}
		&Conv\_D5 &$32\times M\times \widetilde{N}$ &$3\times 3$ &-\\
		\cline{2-5}
		&Conv\_I2 &$32\times M\times \widetilde{N}$ &$3\times 3$ &-\\	
		\cline{2-5}
		&Conv\_I3 &$2\times M\times \widetilde{N}$ &$3\times 3$ &-\\	
		\hline
		\multirow{4}*{CENet}
		&Conv\_C1 &$32\times M\times N$ &$5\times 5$ &-\\
		\cline{2-5}
		&Conv\_D1-Conv\_D4 &$32\times M\times N$ &$3\times 3$ &LReLU\\
		\cline{2-5}
		&Conv\_D5 &$32\times M\times N$ &$3\times 3$ &-\\
		\cline{2-5}
		&Conv\_C2 &$2\times M\times N$ &$3\times 3$ &-\\	
		\hline
	\end{tabular}	
\end{table}

As shown in Fig. \ref{net_flow}, to keep the dimensions of CENet's output consistent with that of $\mathbf{A}$, we expand the last dimension of $\widehat{\widetilde{\mathbf{A}}}$ from $\widetilde{N}$ to $N$ via an operation similar to the generation of $\bar{\mathbf{A}}$ in Section \ref{element_grouping}.
We utilize 2 convolutional layers (``Conv\_C1'' and ``Conv\_C2'' in Fig. \ref{net_flow}) and $N_{CR}$ RBs to construct CENet.
Similar to Conv\_I1 and Conv\_I3 in IENet, Conv\_C1 applies 32 kernels with size 5$\times$5 to extract the features of the input, and Conv\_C2 applies 2 kernels with size 3$\times$3 to match the dimension of the output.
Since the structure of the RBs in CENet is the same as that in IENet, we omit the corresponding description due to space limitation.
Note that $\mathbf{A}=\mathbf{V}+\bar{\mathbf{A}}$; we follow the structure of ResNet and adopt the residual mapping of CENet to learn $\mathbf{V}$, which is then added to the identical mapping of CENet to output the extrapolation result.
Finally, we combine the real and imaginary parts of the resultant in the post-processing module and obtain $\widehat{\mathbf{A}}$.
The structures of IENet and CENet are listed in TABLE \ref{table_Net_structure}.

\subsection{Training Procedure}
\label{training}

Instead of separately training IENet and CENet, we train them simultaneously.
Before training the two networks, we first determine the grouping factor $K$ and then send the pilot signal with length $T$ at the user side.
To generate the dataset, we change the location of the user and collect all the received signal $\mathbf{Y}$ into a set $\mathcal{Y}$ at the AP side.
With the known pilot matrix $\widetilde{\mathbf{X}}$ and $\mathcal{Y}$, we generate the network training dataset as $\mathcal{D}$, where $|\mathcal{D}|=N_{tr}$ is the number of training samples.
One sample in $\mathcal{D}$ contains three matrices and is denoted as $(\widetilde{\mathbf{A}}^{(0)},\widetilde{\mathbf{A}},\mathbf{A})$, where $\widetilde{\mathbf{A}}^{(0)}$ is the initial input, and $\widetilde{\mathbf{A}}$ and $\mathbf{A}$ are the labels of IENet and CENet, respectively.
More details about the dataset generation are presented in Section \ref{simulation}.


During the training, the tunable parameters $\bm{\Theta}$ and $\bm{\Xi}$ are obtained by minimizing the mean square errors (MSEs) between the outputs of the two networks and their corresponding labels.
The loss of IENet is written as
\begin{align}
\label{IENet_loss}
	\mathcal{L}_I=\frac{1}{M_{tr}M\widetilde{N}}\sum_{\mu=1}^{M_{tr}}\left\|[\widetilde{\mathbf{A}}]_{\mu}-[\widehat{\widetilde{\mathbf{A}}}]_{\mu}\right\|_F^2,
\end{align}
where $[\cdot]_{\mu}$ denotes the $\mu$-th sample in the mini-batch and $M_{tr}$ is the batch size for training.

The loss of CENet is written as
\begin{align}
\label{CENet_loss}
	\mathcal{L}_C=\frac{1}{M_{tr}MN}\sum_{\mu=1}^{M_{tr}}\left\|[\mathbf{A}]_{\mu}-[\widehat{\mathbf{A}}]_{\mu}\right\|_F^2.
\end{align}

The above two losses are added together as $\mathcal{L}=\mathcal{L}_C+\rho\mathcal{L}_I$ to jointly update the parameters of IENet and CENet during training, where $\rho$ is the penalty multiplier that evaluates the importance of different losses.
For the networks training, the adaptive moment estimation (Adam) \cite{Adam} algorithm is adopted to achieve the optimal $\bm{\Theta}$ and $\bm{\Xi}$.

\section{Numerical Results}
\label{simulation}

To evaluate the performance of the designed networks, we resort to the indoor scenario `I1' of the DeepMIMO dataset \cite{DeepMIMO}, which is widely used in DL applications for massive MIMO systems.
The BS 10 in the `I1' scenario is adopted as the RIS and is set as a UPA with $8\times 8$ ($N=64$) reflection elements.
The $300$-th user in the $500$-th row is assigned as AP with $M=16$ antennas in the form of a ULA and the users are located in the region from the 1-st row to the 400-th row.
We set the carrier frequency as $f_c=2.4$ GHz.
For both the RIS and AP, the antenna spacing $d$ is set to $\lambda/2$, with $\lambda$ as the carrier wavelength.
The number of the scattering paths for both $\mathbf{h}$ and $\mathbf{G}$ is 5.
To generate the channel dataset, we first obtain $\mathbf{h}$, $\mathbf{G}$ and $\mathbf{A}$ with the location of users and AP.
Next, we determine the group size $K$ and pilot length $T$ to generate $\widetilde{\mathbf{X}}$ and $\mathbf{Y}$.
With $\mathbf{A}$, $\widetilde{\mathbf{X}}$ and $\mathbf{Y}$, we further generate the input-labels tuple $(\widetilde{\mathbf{A}}^{(0)},\widetilde{\mathbf{A}},\mathbf{A})$ as a sample in the dataset.
Since each row in the user's region contains 201 users, the total number of the cascaded channel samples is $80400$.

 \newcommand{\tabincell}[2]{\begin{tabular}{@{}#1@{}}#2\end{tabular}}
 \renewcommand{\arraystretch}{1}
 \begin{table}[!t]
  \centering
  \fontsize{8}{12}\selectfont
  \caption{Performance Comparison of Different Structures under Different Group Sizes.}
  \label{performance_comparison}
  \scalebox{0.9}{
    \begin{tabular}{|c|c|c|c|}
    \hline
    \multirow{3}{*}{\makecell{$K$}}&
    \multicolumn{3}{c|}{Structure}\cr\cline{2-4}
    & \tabincell{c}{IENet+CENet} &\tabincell{c}{IENet+CENet \\ no dense connection in DBs}&\tabincell{c}{Cascaded \\ CNN}\cr
    \hline
    \hline
        {2} & {\bf -31.8776 dB} & -30.3112 dB & -25.5706 dB\cr\hline
        {4} & {\bf -29.9801 dB} & -25.5156 dB & -18.1910 dB\cr\hline
        {8} & {\bf -26.3107 dB} & -23.5888 dB & -17.0309 dB\cr\hline
        {16}& {\bf -22.4233 dB} & -19.4551 dB & -14.3472 dB\cr
    \hline
    \end{tabular}
    }
 \end{table}


We employ 90$\%$ of the dataset for training and the rest for validation.
Specifically, to achieve a trade-off between the network performance and the network training overhead, we adopt $N_{IR}=2$ for IENet and $N_{CR}=4$ for CENet.
For both IENet and CENet, $\beta$ is empirically set as $0.2$.
The total iterations for training is $N_{iter}=300000$ and $\rho=0.1$.
We set the initial learning rate $\eta$ as $1e-3$ and halve it every 50000 iterations.
The Adam optimizer is used for network training with batch size $M_{tr}=16$.
To evaluate the channel extrapolation performance, we use the normalized MSE (NMSE) $\mathbb{E}\left\{\frac{\|\mathbf{A}-\widehat{\mathbf{A}}\|_F^2}{\|\mathbf{A}\|_F^2}\right\}$ as the metric for the designed scheme.


In TABLE \ref{performance_comparison}, we compare the extrapolation performance of three different network structures, i.e., the structure designed in this letter, the structure without the dense connections in DBs and the cascaded CNN.
We set the SNR as 20 dB.
For fairness, the three structures have the same number of layers and the same kernel size at each layer.
It can be seen that the dense connections provide performance gains compared to the structure without the dense connections.
Moreover, both structures with or without the dense connections are superior to the cascaded CNN, which illustrates the effectiveness of the designed structures.

 \begin{figure}
	 \centering
     \includegraphics[width=100mm]{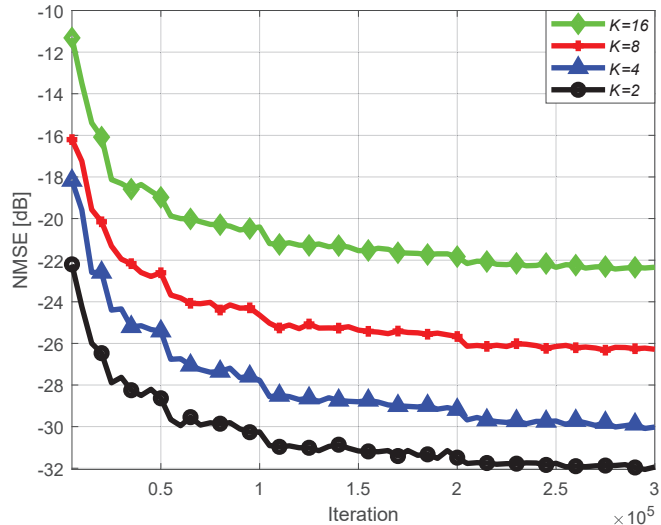}
	 \caption{NMSE of channel extrapolation vs. network training iteration for different $K$ values.}
	 \label{iter_NMSE}
 \end{figure}


Fig. \ref{iter_NMSE} depicts the channel extrapolation performance of the designed jointly training scheme on the validation set versus the network training iterations.
The SNR is set as 20 dB and 4 element group sizes, i.e., $K=2,4,8,16$, are investigated.
It can be seen that the NMSE decreases with the training iterations and almost achieves convergence after 200000 iterations.
Furthermore, Fig. \ref{iter_NMSE} shows that the smaller the group size is, the better the extrapolation performance is; this can be explained, as with a smaller group size, more partial channel information can be obtained, which reduces the difficulty of channel extrapolation.

 \begin{figure}
	\centering
	\includegraphics[width=100mm]{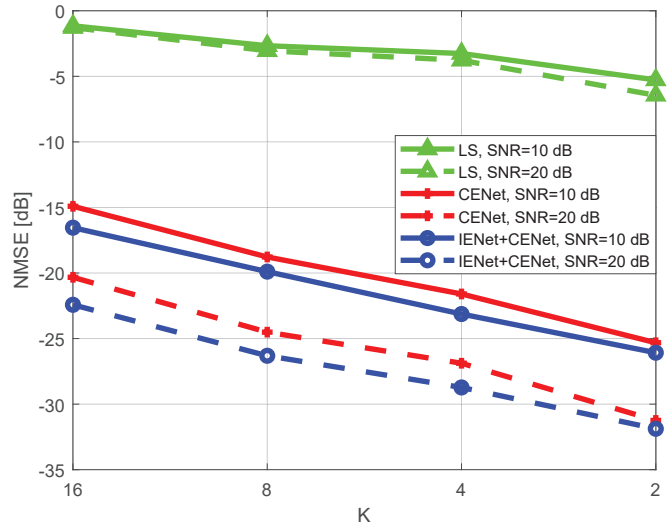}
	\caption{NMSE of channel extrapolation vs. $K$ for different extrapolation schemes.}
	\label{K_NMSE}
 \end{figure}

Fig. \ref{K_NMSE} shows the channel extrapolation performance versus $K$ with SNR=10 dB and 20 dB.
We compare the performance of three different schemes.
Specifically, curves marked with ``LS'' represent the performance that directly expands the LS estimation, i.e., $\widetilde{\mathbf{A}}^{(0)}\otimes\mathbf{1}_K^T$, as the extrapolated channel, curves marked with ``CENet'' represent the performance that adopts $\widetilde{\mathbf{A}}^{(0)}\otimes\mathbf{1}_K^T$ as the input of CENet, and curves marked with ``IENet+CENet'' represent the performance of the designed jointly training scheme.
It can be seen that due to the intra-group interference, the performance of LS estimation is poor.
Both ``CENet'' and ``IENet+CENet'' can provide huge gains compared to LS estimation, which shows the powerful extrapolation ability of CENet.
Moreover, due to the refinement of the partial channels, ``IENet+CENet'' outperforms ``CENet'', which verifies the effectiveness of IENet.

 \begin{figure}
	\centering
	\includegraphics[width=100mm]{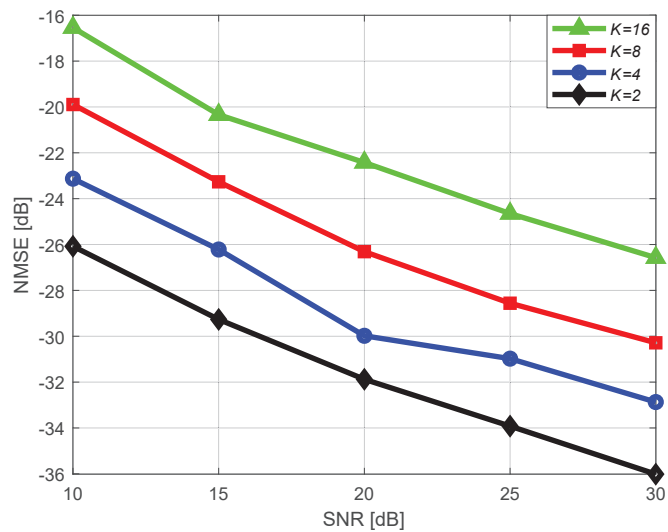}
	\caption{NMSE of channel extrapolation vs. SNR for different $K$ values.}
	\label{SNR_NMSE}
 \end{figure}

In Fig. \ref{SNR_NMSE}, we investigate the channel extrapolation performance of the designed jointly training scheme versus SNR with different $K$ values.
As the SNR increases, the NMSE decreases.
The NMSE with smaller $K$ is always lower than that with larger $K$, which confirms the explanation provided for the result in Fig. \ref{iter_NMSE}.

\section{Conclusion}
\label{conclusion}

In this letter, we have analysed the channel interference caused by the element-grouping strategy and designed two DL-based networks, i.e., IENet and CENet, to obtain the cascaded channels of RIS.
The former aims to obtain the refined partial channel information from the coarse LS estimation, while the latter tries to extrapolate the full channel information from the refined partial one.
To achieve enhanced performance, we have cascaded the two networks and jointly trained them.
Simulation results have illustrated that the proposed scheme provides significant gain compared to the conventional element-grouping method without interference elimination.

\linespread{1.42}
\balance

\end{document}